\documentclass[12pt]{article}
\usepackage{euscript,amsmath, amssymb, amsfonts}
\usepackage{color}

\newcommand{\HH}{\mathcal H }

\newcommand{\HG}{$\mathcal H$}

\newcommand{\tr}{\text{tr}}

\newcommand{\m}{\mathfrak {s} }

\newcommand{\be}{\begin{equation}}
\newcommand{\ee}{\end{equation}}
\newcommand{\bee}{\begin{eqnarray}}
\newcommand{\eee}{\end{eqnarray}}
\newcommand\nn{ \nonumber \\}

\newcounter{theorem}
\newcommand{\theorem}{\par\refstepcounter{theorem}
           {\bf Theorem \arabic{section}.%
           \arabic{theorem}. }}

\makeatletter \@addtoreset{theorem}{section}

\newcounter{corollary}

\makeatletter \@addtoreset{corollary}{section}

\newcounter{lemma}

\makeatletter \@addtoreset{lemma}{section}

\newcounter{proposition}
\newcommand{\proposition}{\par\refstepcounter{theorem}
           {\bf Proposition \arabic{section}.%
           \arabic{theorem}. }}

\makeatletter \@addtoreset{proposition}{section}

\newcounter{conjecture}

\makeatletter \@addtoreset{conjecture}{section}

\newcounter{remark}
\newcommand{\remark}{\par\refstepcounter{theorem}
           {\bf Remark \arabic{section}.%
           \arabic{theorem}. }}

\makeatletter \@addtoreset{remark}{section}

\newcounter{definition}
\newcommand{\definition}{\par\refstepcounter{theorem}
           {\bf Definition \arabic{section}.%
           \arabic{theorem}. }}

\makeatletter \@addtoreset{definition}{section}

\newenvironment{proof}[1][Proof]{\noindent\textsf{#1.\ }}
{\hfill {\small $\square$}}

\makeatletter \@addtoreset{equation}{subsection}

\begin{document}

\begin{flushright}
FIAN-TD-2021-17  \hspace{2.6cm}{}~\\
arXiv: yymm.nnnnnn [hep-th] \hspace{0.5cm}{}~
\end{flushright}

\vspace{1cm}

\begin{center}

{\Large \bf Ideals generated by traces in the symplectic reflection

\medskip
algebra $H_{1,\nu_1, \nu_2}(I_2(2m))$. II.}

\vspace{2.5cm}

I.A. Batalin,\footnote{E-mail: batalin@lpi.ru}
\fbox{S.E. Konstein}\,, and %\footnote{E-mail: konstein@lpi.ru}
I.V. Tyutin\footnote{E-mail: tyutin@lpi.ru}

\hspace{2cm}

{\it I.E. Tamm Department of Theoretical Physics, P.N. Lebedev Physical Institute, RAS 119991, Leninsky prosp., 53, Moscow, Russia.}

\vspace{3.5cm}

Abstract

\end{center}

The associative algebra of symplectic reflections $\mathcal H:= H_{1,\nu_1, \nu_2}(I_2(2m))$ based on the group generated by the root system $I_2(2m)$ has two parameters, $\nu_1$ and $\nu_2$. For every value of these parameters, the algebra $\mathcal H$ has an $m$-dimensional space of traces. A~given trace $\tr$ is called degenerate if the associated bilinear form $B_{\tr}(x,y)=\tr(xy)$ is degenerate. Previously, there were found all values of $\nu_1$ and $\nu_2$ for which there are degenerate traces in the space of traces, and consequently the algebra $\mathcal H$ has a two-sided ideal. We proved earlier that any linear combination of degenerate traces is a degenerate trace. It turns out that for certain values of parameters $\nu_1$ and $\nu_2$, degenerate traces span a 2-dimensional space. We prove that non-zero traces in this $2d$ space generate three proper ideals of $\mathcal H$.

\newpage

\section{Introduction}
This paper is a continuation of \cite{tmf}; we advise the reader to recall \cite{tmf}.

\section{The associative algebra $H_{1,\nu_1,\nu_2}(I_2(n))$ with $n$ even}\label{App}

\subsection{The group $I_2(2m)$}

\definition
{We denote the finite subgroup of $O(2,\mathbb R)$, generated by
the root system $I_2(2m)$, also by $I_2(2m)$.
It consists of
$2m$ reflections $R_k$
and $2m$ rotations $S_k:=R_k R_0$, where $S_0=S_{2m}$ is the unit in $I_2(2m)$.
% and   $S_m^2=S_0$.     %\SK{the definition needed!}.
The indices of $S_k$ and $R_k$ belong to $\mathbb Z/n \mathbb Z$.
These elements satisfy the relations
\begin{equation}\label{ARR}     % %\nonumber
R_k R_l = S_{k-l},\qquad
S_k S_l = S_{k+l},\qquad
R_k S_l = R_{k-l},\qquad
S_k R_l = R_{k+l}.
\end{equation}
}

Evidently, the $R_{2k}$ belong to one conjugacy class and the $R_{2k+1}$ belong to another class.
The rotations $S_k$ and $S_l$ constitute a conjugacy class if $k+l=2m$.

It is convenient to use the following basis $L_p$, $Q_p$ instead of $R_k$ and $S_k$ in the group algebra $\mathbb C[I_2(2m)]$:

\definition
\begin{eqnarray}\label{ALQbas}
&& L_p:=\frac 1 {2m} \sum_{k=0}^{2m-1} \lambda^{kp} R_k, \qquad 
Q_p:=\frac 1 {2m} \sum_{k=0}^{2m-1} \lambda^{-kp} S_k,
\nonumber\\
&& \lambda=\exp\left(\frac {\pi i} m \right),\qquad  p=0,\,...\,2m-1.
\end{eqnarray}

\subsection{Symplectic reflection algebra
$H_{1,\nu_1, \nu_2}(I_2(2m))$}\label{Asrai2}

 For a general definition of Symplectic reflection algebra, see, e.g.,
\cite{sra}.
Here we restrict ourselves to $H_{1,\nu_1, \nu_2}(I_2(2m))$ only, introducing,
for convenience, the new parameters
\begin{equation}\label{A3.5}
\mu_0:=m ( {\nu_1+\nu_2}) ,\ \ \ \mu_1 :=m ( {\nu_1-\nu_2}).
\end{equation}
%%%%%%%%%%%%%%%%%%%%%%%%%%%%%%%%%%%%%%%%%%%%%%%%%%%%%%%%%%%%%%%%%%%%%%%%%%%%%
\iffalse
Let
\be
\delta_i=1,\ \mbox{if}\ i=0,\ \mbox{and} \ \delta_i=0,\ \mbox{if}\ i\ne 0.
\ee
\fi
%%%%%%%%%%%%%%%%%%%%%%%%%%%%%%%%%%%%%%%%%%%%%%%%%%%%%%%%%%%%%%%%%%%%%%%%%%%%%

Here and below we use the following notation. Let $I$ be a logical expression; set
\be
\delta_{I}:=\left\{%
\begin{array}{c}
1, \ \mbox{if  $I$ is true};\\
0, \ \mbox{if $I$ is false};
\end{array}
%\ \ \SK{\ kak\ by\ eto\ napisatx\ akkuratnee?}
\right.
\ee
\begin{equation*}
\end{equation*}
e.g. $\delta_{i=j}=\delta_{i\,j}$ (the Kronecker delta), $\delta_{p\ge q}=1$ if $p\ge q$ and $\delta_{p\ge q}=0$ if $p < q$.

\definition
{The symplectic reflection algebra
$\mathcal H :=H_{1,\nu_1, \nu_2}(I_2(2m))$} is an associative algebra
of polynomials in $a^{\alpha},b^{\alpha}$, where $\alpha=0,1$,
with coefficients in $\mathbb C [I_2(2m)]$, satisfying the relations%
\begin{eqnarray}
% G-H
\label{ARH}     % %\nonumber
R_k a^\alpha=  - {\lambda }^k b^\alpha R_k ,&\qquad&
R_k b^\alpha = - {\lambda }^{-k} a^\alpha R_k , \\
\label{ASH}     % %\nonumber
S_k a^\alpha =    {\lambda }^{-k} a^\alpha S_k ,&\qquad&
S_k b^\alpha =   {\lambda }^k     b^\alpha S_k ,        \nonumber
\end{eqnarray}
and
% H-H
\begin{eqnarray}\label{AHH2}
&\left[a^\alpha,\,b^\beta\right]=\varepsilon^{\alpha\beta}
\left( 1+ \mu_0  L_0 +  \mu_1   L_{m}\right), &\\
&\left[a^\alpha,\,a^\beta\right]=\varepsilon^{\alpha\beta}
\left(  \mu_0   L_1 +  \mu_1  L_{m+1} \right),&\nn
&\left[b^\alpha,\,b^\beta\right]=\varepsilon^{\alpha\beta}
\left(  \mu_0   L_{-1} +  \mu_1  L_{m-1} \right),&
      \nonumber
     \end{eqnarray}
where $\varepsilon^{\alpha\beta}$ is the skew-symmetric tensor with $\varepsilon^{0\,1}=1$.

The relations (\ref{ARH})
imply
\begin{eqnarray}
\label{ALH}     % %\nonumber
&L_p a^\alpha = -   b^\alpha L_{p+1} ,\qquad
&L_p b^\alpha = -   a^\alpha L_{p-1} ,\\
\label{AQH}
&Q_p a^\alpha = a^\alpha Q_{p+1} ,\qquad
&Q_p b^\alpha = b^\alpha Q_{p-1} ,\\
% G-G
\label{AGG}
&L_k L_l =  \delta_{k=-l} Q_l, \qquad &L_k Q_l =  \delta_{k=l} L_l,\\
&Q_k L_l =  \delta_{k=-l} L_l, \qquad & Q_k Q_l =  \delta_{k=l} Q_l,
%\\   &\ \ \mbox{where }\delta_{k=0}\defeq \delta_{k0},      %\nonumber
\end{eqnarray}

\subsection{Subalgebra of singlets}

The algebra \HG contains
the Lie subalgebra  $sl_2$ of inner derivations with the generating elements
\be
T^{\alpha\beta}:=\frac 1 2 (\{a^\alpha,\, b^\beta \}+\{b^\alpha,\, a^\beta \})
\ee
which
act on \HG \ as follows
\begin{equation}      %\nonumber
 f\mapsto \left[f,T^{\alpha\beta}\right] \ \
\text{for each } f\in\HH.
\end{equation}
%\SK{ }
We say that the element $f\in\HH$ is a~{\it singlet} if $\left[f,T^{\alpha\beta}\right]=0$ for each $\alpha,\,\beta$
and denote the subalgebra
consisting of all the
singlets in $\HH$  by $\HH_0$.

Let the skew-symmetric tensor $\varepsilon_{\alpha\beta}$ be such that
$\varepsilon_{0\,1}=1$ and $\varepsilon_{\alpha \beta} \varepsilon^{\gamma \beta}=\delta_{\alpha}^{\gamma}$.
Set
\begin{eqnarray}      %\nonumber
\m:=\sum_{\alpha,\beta=0,1}
\frac 1 {4i} (\{a^\alpha,\, b^\beta \}-\{b^\alpha,\, a^\beta \})\varepsilon_{\alpha\beta}.
\end{eqnarray}

Then
\begin{eqnarray}      %\nonumber
&&   [\m ,\,Q_p] = [\m ,\,S_k] = [T^{\alpha\beta},\,\m]=0,     \\
&&   \m L_p =-L_p\m, \ \m R_k =-R_k\m,                         \\
&& (\m -i(\mu_0 L_0 + \mu_1 L_m) )a^\alpha = a^\alpha (\m + i+i (\mu_0 L_0 + \mu_1 L_m)).        %\nonumber
\end{eqnarray}

\proposition
\cite{tmf}
{\it
If $f\in\HH_0$, then $f$ has the form
\begin{equation*}
f=\sum_{p=0}^{2m-1}(\phi_p Q_p+\psi_p L_p),
\mbox{\ \ where\ \ } \phi_p,\,\psi_p \in \mathbb C[\m].
\end{equation*}
}

%%%%%%%%%%%%%%%%%%%%%%%%

\section{Ideals generated by degenerate traces
%\SK{}
}
%%%%%%%%%%%%%%%%%%%%%%%%

We call the trace $tr$ degenerate if symmetric invariant bilinear form $B_{tr}: \ B_{tr}(x,\,y)=tr(xy)$
is degenerate.

It is clear that the kernel of $B_{tr}$ is an ideal in \HG, we will denote it $\mathcal I_{tr}$.

It is shown in \cite{KT19} that for any degenerate trace $tr$, the ideal $\mathcal I_{tr}$
is completely determined by $\mathcal I_{tr}\bigcap \HH_0$.

\subsection{The values of the trace on $\mathbb C[I_2(2m)]$}\label{AGAi2}

According to the general result of \cite{KT}, the restriction of the trace to $\mathbb C[I_2(2m)]$
is completely defined by the  value of this trace on the conjugacy classes without the eigenvalue +1 in their spectra.

The group $I_2(2m)$ has the following $m$ conjugacy classes without the eigenvalue +1 in their spectra:
$m-1$ classes consisting of two elements, %\SK{ }
\begin{equation}      %\nonumber
\{S_{p},S_{n-p}\}, \text { where } p=1,...,m-1,
\end{equation}
and one class consisting of 1 element,  %\SK{}
\be
\{S_{m}\}.
\ee

The values of the trace on these conjugacy classes
\begin{equation}\label{A1.2.3}
%\SK{  }
s_k := \tr(S_k)=\tr(S_{n-k}),\ \ %\text{ where }
s_{2m-k}=s_k, \ \ k=1,...,m,
\end{equation}
completely define the trace on \HG, and therefore the dimension of the space of traces
is  equal to $m$.

Besides, the group $I_2(2m)$ has two conjugacy classes each having one eigenvalue +1 in its spectrum:
\begin{equation}      %\nonumber
\{R_{2l} \mid l=0,...,m-1\},\;\{R_{2l+1} \mid  l=0,...,m-1\},
\end{equation}
and one conjugacy class with two eigenvalues +1 in its spectrum: $\{S_0\}$.

The traces on these conjugacy classes (see \cite{tmf})
are equal to
\begin{eqnarray}      %\nonumber
\tr(R_{2l})&=&-2\nu _{2}X_1 -2\nu _{1} X_2,
\ (l=0,1,...,m-1),\\
\tr(R_{2l+1})&=&-2\nu _{1} X_1 - 2\nu _{2} X_2 ,
\ (l=0,1,...,m-1), \\
% &\ \ & l=0,1,...,m-1,      %\nonumber
%  \\
%
\label{As0}
\tr(S_{0}) &=& 2(\nu _{1}^{2}+\nu _{2}^{2})m
X_1+4\nu _{1}\nu_{2} m X_2,
\end{eqnarray}
where
\begin{eqnarray}      %\nonumber
 X_1:= \sum_{l=1}^{m-1}s_{2l}\sin ^{2}\left(\frac{\pi
l}{m}\right),
\qquad
 X_2:=\sum_{l=0}^{m-1}s_{2l+1}\sin ^{2}%
\left(\frac{\pi (2l+1)}{2m}\right).      %\nonumber
\end{eqnarray}

We note also that
\begin{eqnarray}      %\nonumber
&& \tr(L_0)=-\frac {\mu_0} m (X_1+X_2),\quad \tr(L_m)=-\frac {\mu_1} m (X_1-X_2),\quad
\tr (L_p)=0 \text{ for } p\ne0,\,m,
\\
&& \tr(S_0)=-\mu_0 \tr(L_0)-\mu_1 \tr(L_m). \ %\SK{\ srawni\ s\ (2.3.6)}
      %\nonumber
\end{eqnarray}

%%%%%%%%%%%%%%%%%%%%%%%%
%%%%%%%%%%%%%%%%%%%%%%%%

\subsection{Generating functions of the trace}\label{Agenfuni2}

Set $\mathcal L :=\mu_0 L_0 + \mu_1 L_m$.

For each trace $\tr$, we define the following set of generating functions on \HG:
\begin{eqnarray}\label{A1.33}
&& F_p(t):=\tr(\exp(t(\m-i \mathcal L ))Q_p),
  \\
&& \Psi_p(t):=\tr(\exp(t\m)L_p),      %\nonumber
\end{eqnarray}
where $p=0,...,2m-1$.
From $\m L_p = - L_p \m$ and definition of the trace it follows  that
\begin{equation}      %\nonumber
\Psi_p(t)=\Psi_p(0).
\end{equation}
%
%  Since $L_0 Q_p=0$ for $p\ne 0$ and $L_m Q_p=0$ for $p\ne m$,
%  definition (\ref{A1.33}) implies that
%  \begin{eqnarray} %\nonumber
%  F_p(t)&=& tr(\exp(t \m)Q_p) \text{ for } p\ne 0,m,
%  \nn
%  F_0(t)&=&tr(\exp(t(\m-i\mu_0 L_0))Q_0),
%  \nn
%  F_m(t)&=&tr(\exp(t(\m-i\mu_1 L_m))Q_m). %\nonumber
%  \end{eqnarray}
%
We also consider the functions $\Phi_p(t):=\tr(\exp(t(\m+i \mathcal L))Q_p)$
related with the functions $F_p$
by the formula
\begin{equation}      %\nonumber
\Phi_p(t)=F_p(t)+2i\Delta_p(t), \\
\text{where } \Delta_p(t)=\delta_p \sin(\mu_0 t)\tr(L_0)+\delta_{m-p} \sin(\mu_1 t)\tr(L_m).
      %\nonumber
\end{equation}

In \cite{tmf} we derived the following system of equations
\begin{eqnarray}\label{Aeqgenfun}
\frac d {dt} F_p - e^{it}\frac d {dt}F_{p+1}=iF_p +
ie^{it}F_{p+1}+2i\frac d {dt}\left(e^{it}\Delta_{p+1} \right).
\end{eqnarray}
%
%%%%%%%%%%%%%%%%%%%%%%%%%%%%%%%%%%%%%%%%%%%%%%%%%
%%%%%%%%%%%%%%%%%%%%%%%%%%%%%%%%%%%%%%%%%%%%%%%%%
%
Next, consider the Fourier transform of (\ref{Aeqgenfun}),
%namely, we consider
\begin{eqnarray}      %\nonumber
&& G_k:= \sum_{p=0}^{2m-1} \lambda^{kp}F_p, \text{ where } k=0,...,2m-1,
\label{Afur}
\\
&& \widetilde{\Delta}_k:=\sum_{p=0}^{2m-1} \lambda^{kp}\Delta_{p+1}=
\lambda^{-k}\left( \sin(\mu_0 t)\tr(L_0) + \lambda^{km}\sin(\mu_1 t)\tr(L_m)    \right),
 \nn
&& \qquad\qquad\qquad \text{ where } k=0,...,2m-1 \qquad
 \text{ and $ \lambda:=e^{i\pi/m}$,}
      %\nonumber
\end{eqnarray}
%and obtain the system of equation
which satisfies the equation
\begin{equation}      %\nonumber
\frac d {dt} G_k = i\frac {\lambda^k+e^{it}}{\lambda^k-e^{it}} G_k +
\frac {2i\lambda^k}{\lambda^k-e^{it}} \frac d {dt}\left(e^{it}\widetilde{\Delta}_k\right)
\end{equation}
with initial conditions
\begin{equation}     \label{Ainit}
G_k(0)=s_k, \text{ where } k=0,...,2m-1, \text{ \ \ \ and\ \ \ }  s_k=s_{2m-k}
\end{equation}
and where the $s_k$ are defined by Eq. (\ref{A1.2.3}) for $k=1,\,...,\,2m-1$,  and
$s_0 := \tr(S_0)$ is defined by Eq. (\ref{As0}).
The value $s_0$ depends linearly on
$s_k$, where $k=1,...,m$  (see Eq. (\ref{As0}) and take in account the relations $s_k=s_{2m-k}$).

The solution of the equations for $G_k$ has the form:
\begin{eqnarray}\label{AGK}
G_k(t)= \frac {e^{it}f_k(t)} {(e^{it}-\lambda^k)^2},
\end{eqnarray}
where
\begin{eqnarray}\label{A1.47}
\!\!\!\!\!\!\!\!\!\! f_k(t)&=&\frac{2\lambda ^{k}}{m}X_{+}[1-\cos (t\mu _{0})]+(-1)^{k}\frac{2\lambda
^{k}}{m}X_{-}[1-\cos (t\mu _{1})]+
\nn
&+&\frac{2i}{m}(e^{it}-\lambda ^{k})[\mu _{0}X_{+}\sin (t\mu
_{0})+(-1)^{k}\mu _{1}X_{-}\sin (t\mu _{1})]-4 \lambda^k s^\prime_k.
%+(1-\lambda ^{k})^{2}s_{k},
\end{eqnarray}
Here
\bee
&&X_{\pm}:=X_1 \pm X_2,   \\
%\begin{equation}      %\nonumber
&&s^{\prime}_k := s_{k} \sin^2\left(\frac {\pi k} {2m} \right),
\quad k=1,...,2m-1, \qquad s^\prime_0=0.
\eee

Note that the functions $G_k$ in Eq (\ref{AGK}) do not depend on the signs of $\mu_0$ and $\mu_1$.

So we can consider $\mu_0$ and $\mu_1$  positive if they are non-zero.

\iffalse
Note
\begin{equation}
(1-\lambda ^{k})^{2}s_{k}=-4 \lambda^k \sin^2 \frac {k\pi}{2m} s_k = -4 \lambda^k s^\prime_k
\end{equation}
\fi
%%%%%%%%%%%%%%%%%%%%%%%%

\subsection{The degeneracy conditions for the trace}\label{Adegcond}

It is proved in \cite{tmf} that the trace $\tr$ is degenerate if and only if
the functions $F_p$ (and, as a consequence, $G_k$) have the form $\sum_{i=0}^{n_p}\exp(t\omega_i)q_i(t)$,
where the $q_i$ are polynomials. Further, these conditions yield that the trace $\tr$ is degenerate if and only if
% Laurent polynomials in $z:= \exp(it)$
%with coefficients in $\mathbb C [t]$.
%
%Then the system of linear equations for
the parameters $s^\prime_k$ satisfy the following system of linear equations
\begin{eqnarray}\label{Aur1}
&& \!\!\!\!\!\!\!\!\!\!\!\!\!\!\!\!\!\!\!\!
\left(1 - \cos \left(\frac {\pi}{ m} k\mu _{0}\right)\right)X_{+}
+ (-1)^{k} \left(1 - \cos \left(\frac {\pi} {m} k\mu _{1}\right)\right)X_{-}
=2 m  s^\prime_{k},  \quad k=1,...,2m-1,\quad    %\SK{Aur1}
\\
&&  \!\!\!\!\!\!\!\!\!\!\!\!\!\!\!\!\!\!\!\!
s^\prime_{2m-r}=s^\prime_r, \quad r=1,...,m, \label{Aur2} \\
\label{Aur3}
&& \!\!\!\!\!\!\!\!\!\!\!\!\!\!\!\!\!\!\!\!
X_{\pm}=X_1 \pm X_2,\\
\label{Aur4}
&& \!\!\!\!\!\!\!\!\!\!\!\!\!\!\!\!\!\!\!\!
 X_1= \sum_{1\le l\le m-1} s^\prime_{2l},               \\
\label{Aur5}
&&  \!\!\!\!\!\!\!\!\!\!\!\!\!\!\!\!\!\!\!\!
X_2= \sum_{0\le l\le m-1}s^\prime_{2l+1},
\end{eqnarray}
and the parameters $\mu_0$ and $\mu_1$ are defined from the condition
that this system has a nonzero solution.

\iffalse
$$
-\frac{2}{m}\lambda^k\left(1 - \cos \left(\frac {\pi}{ m} k\mu _{0}\right)\right)X_{+}
-\frac{2}{m}\lambda^k (-1)^{k} \left(1 - \cos \left(\frac {\pi} {m} k\mu _{1}\right)\right)X_{-}
=-4 \lambda^k s^\prime_{k},  \quad k=1,...,2m-1,
$$
\fi

%%%%%%%%%%%%%%%%%%%%%%%%

\theorem\label{Ath1}
{\it Let $m \geqslant 2$. Then
the system of equations \textup{(\ref{Aur1})-(\ref{Aur5})} has nonzero solutions
at the following values of the parameters $\mu_0$ and $\mu_1$ only:
\begin{eqnarray}\label{Ak1}
&& \mu_0 \in \mathbb Z\diagdown m\mathbb Z, \qquad \mu_1 \in \mathbb Z\diagdown m\mathbb Z,
\\
\label{Ak3}
&& \mu_0 \in \mathbb Z\diagdown m\mathbb Z, \qquad \text{any } \mu_1,
\\
\label{Ak4}
&& \mu_1 \in \mathbb Z\diagdown m\mathbb Z, \qquad \text{any } \mu_0,
\\
\label{Ak5}
&& \mu_0=\pm \mu_1+ m(2l+1), \qquad l=0,\,\pm 1,\, \pm 2,\,...
\end{eqnarray}
Here,

1. In  case \textup{(\ref{Ak1})}, the system of equations  \textup{(\ref{Aur1})-(\ref{Aur5})} has a two-parameter family
of solutions;

2.
In case  \textup{(\ref{Ak3})}, if $\mu_1 \notin \mathbb Z \diagdown m\mathbb Z$,
then the system of equations  \textup{(\ref{Aur1})-(\ref{Aur5})} has a one-parameter family of solutions
with $X_-=0$,

3.
In case  \textup{(\ref{Ak4})}, if $\mu_0 \notin \mathbb Z \diagdown m\mathbb Z$,
then the system of equations  \textup{(\ref{Aur1})-(\ref{Aur5})} has a one-parameter family of solutions
with $X_+=0$,

4.
In case  \textup{(\ref{Ak5})}, if $\mu_0, \mu_1 \notin \mathbb Z \diagdown m\mathbb Z$,
then the system of equations  \textup{(\ref{Aur1})-(\ref{Aur5})} has a one-parameter family of solutions
with $X_1=0$.
}

\iffalse
\remark
Theorem
%\ref{Ath1}
A.5
is proved for $m\geqslant 2$,
nevertheless it describes also the case $m=1$ correctly.

If $m=1$ then the cases (\ref{Ak1}) -- (\ref{Ak4})  disappear,
and the case (\ref{Ak5}) shows that
\begin{equation}\label{Arem}
\text{at least one of $\nu_1$ and $\nu_2$
is half-integer.}
\end{equation}
Because
$H_{1,\nu_1, \nu_2}(I_2(2)) \simeq H_{1,\nu_1}(A_1)\otimes H_{1,\nu_2}(A_1)$,
the statement (\ref{Arem}) follows also from \cite{V}, where
the singular values of $\nu$ and ideals in $H_{1,\nu}(A_1)$
were found.
\fi
%%%%%%%%%%%%%%%%%%%%%%%%%%%%%%%%%

\section{Explicit expressions for generating functions $F_p$.}

Consider the associative algebra \HG \, with parameters $\nu_1,\,\nu_2$ such that
%\SK{ gde mu v nu opredeleno ranee!}
\[
\mu_0 \in \mathbb Z\diagdown m\mathbb Z, \qquad \mu_1 \in \mathbb Z\diagdown m\mathbb Z
\]
and find explicit expressions for generating functions $F_p$.

\subsection{Explicit expressions for generating functions $G_k$ of degenerate trace.}

Recall, that
\be
G_k=\sum_{p=0}^{n-1} \lambda^{kp}F_p, \qquad   F_p=\frac 1 n \sum_{k=0}^{n-1} \lambda^{-kp}G_k.
\ee

Introduce the notation
\be
z:=e^{it}, \qquad  z_k:=z\lambda^{-k} \ \ (k=0,...,n-1),
\ee
and substitute the expression (\ref{Aur1}) into the formula (\ref{AGK}). Since
${\mu_0 \in \mathbb Z\setminus m\mathbb Z}$ and ${\mu_1 \in \mathbb Z\setminus m\mathbb Z}$,
we can introduce the polynomials $P_{\mu_0}$ and $P_{\mu_1}$ of variable $x$ by the formula
\begin{equation*}
P_{\mu}(x):=\frac{x}{(1-x)^{2}}\left( \mu (x-1)-(x^{\mu }-1)\right).
\end{equation*}%
The polynomial $P_{\mu}$ is a polynomial of $x$ and can be represented in the form
\begin{equation*}
P_{\mu}(x)=\begin{cases}
0&\text{if~}\mu =1, \\
-\sum_{q=\mu -1}^{1}(\mu -q)x^{q}&\text{if~}\mu >1 .
\end{cases}%
\end{equation*}

Now we can notice that the functions $G_k$ can be expressed in the form
\be
G_k(t)=G_k^{(+)}(t)X_+ + G_k^{(-)}(t)X_-,
\ee
where
\bee
G_k^{(+)}(t)&=& -\mu_0 z^{\mu_0 }+z^{\mu_0}P_{\mu_0}(z^{-1}\lambda ^{k})
             + z^{-\mu_0 }P_{\mu_0}(z\lambda^{-k}) ,   \\
G_k^{(-)}(t)&=& -(-1)^k\mu_1 z^{\mu_1 }+(-1)^k z^{\mu_1}P_{\mu_1}(z^{-1}\lambda ^{k})
             + (-1)^k z^{-\mu_1 }P_{\mu_1}(z\lambda^{-k}),\\
             &&  \nonumber
\eee
or
\bee
G_k^{(+)}(t)&=& \frac{1}{i}\frac{d}{dt}%
\sum_{q=-\mu_0 }^{-1}z^{-q-\mu_0 }\lambda ^{kq}
-\frac{1}{i}\frac{d}{dt}\left( z^{\mu_0 }+\sum_{q=1}^{q=\mu_0
-1}z^{\mu_0 -q}\lambda ^{kq}\right),
                       \\
G_k^{(-)}(t)&=& (-1)^k \frac{1}{i}\frac{d}{dt}%
\sum_{q=-\mu_1 }^{-1}z^{-q-\mu_1 }\lambda ^{kq}
- (-1)^k \frac{1}{i}\frac{d}{dt}\left( z^{\mu_1 }+\sum_{q=1}^{q=\mu_1
-1}z^{\mu_1 -q}\lambda ^{kq}\right).
                       \\
             && \nonumber
\eee

%%%%%%%%%%%%%%%%%%%%%%%%%%%%%%%%%%%%%%%%%%%%%%%%%%%%%%%%%%%
Now we can write down the functions $F_p(t)$ in the form 
\be
F_p(t)=F^{(+)}_p(t)X_+ + F^{(-)}_p(t)X_-,
\ee
where
\be
F^{(\pm)}_p(t)=\frac 1 n \sum_{k=0}^{n-1}\lambda^{-kp}G^{(\pm)}_k.
\ee

%%%%%%%%%%%%%%%%%%%%%%%%%%%%%%%%%%%%%%%%%%%%%%%%%%%%%%%%%%%
%%%           F_p
%%%%%%%%%%%%%%%%%%%%%%%%%%%%%%%%
First, %\DL{}%Before do it,
introduce the values
\begin{eqnarray}
d_{\alpha} &=&\left[ \frac{\mu_{\alpha} }{n}\right] ,\mbox{ \ where\  $\alpha=0,1$, and $[x]$ is an integer part of $x$}, \\
\widetilde{\mu_{\alpha} } &:&=\mu_{\alpha} -nd_{\alpha},\text{ \ \ \ \ }1\leqslant \widetilde{\mu_{\alpha} }%
\leqslant n-1.
\end{eqnarray}%
%%%%%%%%
Each value $q\in \{-\mu_{\alpha} ,...,\mu_{\alpha}  \}$ can be represented in the form%
\begin{equation*}
q=sn+q^{\prime }
\end{equation*}%
where
%$-d_{\alpha}-1\leqslant s\leqslant d_{\alpha},$ \ \ $0\leqslant q^{\prime }\leqslant n-1. $
%
%
\bee
\mbox{\ }s &\in&\{-d_{\alpha} -1,...,d_{\alpha}\} , \\
q^{\prime } &\in&\{0,...,n-1\}.
\end{eqnarray}%
%%%%%%%%%%%%%%%%%%%%%%%%%%%%%%%%%%%%%%%%
Now we can write
\bee
F_{p}^{(+)}(z)& =&\frac{1}{i}\frac{d}{dt}\left(
\sum_{s=-d_{ 0 }}^{s=-1}z^{-\mu_{ 0 } -sn-p}\delta _{\mu_{ 0 } >n}+z^{-\mu_{ 0 } +(d_{ 0 }+1)n-p}\mbox{ }%
\delta _{n-1\geqslant p\geqslant n-\widetilde{\mu_{ 0 } }}\right)-          \nonumber \\
&-&\frac{1}{i}\frac{d}{dt}\left(
\sum_{s=0}^{d_{ 0 }}z^{\mu_{ 0 } -sn}\delta _{p=0}+\sum_{s=0}^{d_{ 0 }-1}z^{\mu_{ 0 } -sn-p}\delta
_{p\neq 0}\delta _{\mu_{ 0 } >n}+z^{\mu_{ 0 } -d_{ 0 }n-p}\delta _{1\leqslant p\leqslant
\widetilde{\mu_{ 0 } }}\right).
\nonumber\\
\eee
Further,
%%%%%%%%%%%%%%%%%%%%%%%%%%%
%%%%%%%
\bee
F_{p}^{(-)}
(z)  &= &\frac{1}{i}\frac{d}{dt}%
\left( \sum_{s=-d_{1}}^{s=-1}z^{-\mu _{1}-sn- \rho(p) }\delta _{\mu
_{1}>n}+z^{-\mu _{1}+(d_{1}+1)n- \rho(p) }\mbox{ }\delta _{n-1\geqslant
 \rho(p) \geqslant n-\widetilde{\mu _{1}}}\right)         \nonumber \\
%%%
%%%
&&
\!\!\!\!\!\!\!\!\!\!\!\!\!\!\!\!\!\!\!\!\!\!\!\!
-\frac{1}{i}\frac{d}{dt}%
\left( \sum_{s=0}^{d_{1}}z^{\mu _{1}-sn}\delta _{ \rho(p)
=0}+\sum_{s=0}^{d_{1}-1}z^{\mu _{1}-sn- \rho(p) }\delta _{ \rho(p) \neq
0}\delta _{\mu _{1}>n}+z^{\mu _{1}-d_{1}n- \rho(p) }\delta _{1\leqslant
 \rho(p) \leqslant \widetilde{\mu }_{1}}\right),
\nonumber\\
%
%%%%%%%%%%%%%%%%%%%%%%%%%%%%%%%%%%
%%%%%%%%%%%%%%%%%%%%%%
\eee%
where
\be
\rho(p)=
\begin{cases}
p+m&\text{if \ \ }0\leqslant p<m, \\
p-m&\text{ if \ \ }m\leqslant p<2m,
\end{cases}
\ee
it is clear that $\rho(\rho(p))=p$.

\proposition\label{Fpm2} %\SK{ proweritx !!!}
The functions $F_{p}^{(\pm)}$ have the following form
\be\label{Fform}
F_{p}^{(\pm)}=i\frac{d}{dt}\sum_{\ell=-\ell_{\pm}}^{\ell_{\pm}} a_{p,\,\ell}^{\pm}z^{\ell},
\ee
where $\ell_{+}=\mu_0$, $\ell_{-}=\mu_1$, and the $a_{\,p,\,\ell}^{\pm}$ are equal to either $sgn (\ell)$ or $0$.

\proposition\label{Fpm2} %\SK{ proweritx !!!}
If there exist $p\ne 0,\,m$ and $\ell$  such that $a_{p,\,\ell}^{+}\ne 0$
and $a_{p,\,\ell}^{-}\ne 0$, then there exists an odd integer $o$ such that
$\mu_0-\mu_1=om$.

\subsection{Generating functions of the trace II}\label{Agenfuni2-II}

Now define one more set of generating functions of the trace, connected with Eq(\ref{A1.33}):
\begin{eqnarray}\label{A1.33}
 \widetilde{F}_p(t):=\tr(\exp(t \m)Q_p),
\end{eqnarray}
where $p=0,...,2m-1$.

Evidently, $\widetilde{F}_p(t)={F}_p(t)$ if $p\ne 0,\, m$
since
$\mathcal L Q_p=0$ if $p\ne 0,\, m$.

Let $\tr$ be a~degenerate trace. Then, it is possible to express $\widetilde{F}_p$ via ${F}_p$ for $p=0,\,m$.
%%%%%%%%%%%%%%%%%%%%%%%%%%%%%%%%%%%%%%%%%%%%%%%%%%%%%%%%%%%%%%%%%%%%

\subsection{The generating functions $\widetilde F^{}_0=\tr\left(\exp(t\m )Q_0 \right)$ \\ and
$\widetilde F^{}_m =\tr\left(\exp(t\m )Q_m \right)$
  for the degenerate trace}

 Let $\mu_0, \, \mu_1 \in\mathbb Z \setminus m\mathbb Z$.
Express $\widetilde{F}_p$ via ${F}_p$ for $p=0,\,m$.

\iffalse
In this section we introduce the function
\be
%\label{F}
           \nonumber
\widetilde F^{\tr}_0:=\tr\left(\exp(t \frak s )Q_0 \right)
\ee
and express it via $F_0^{\tr}$.
\fi

{\proposition\label{p121} Let $p=0,\,m$, then
$\widetilde F^{}_p$ is an even function of $t$:
\be\label{121}
\widetilde F^{}_p=\tr(\cosh(t\m)Q_p) .
\ee
}

Indeed, $\widetilde F^{}_p=\tr(\cosh(t\m)Q_p+\sinh(t\m)Q_p)$ and
$\tr(\sinh(t\m)Q_p)=0$ since
\bee
&& \tr(\sinh(t\m)Q_p)=\tr((\sinh(t\m)L_p) L_p)=\tr(L_p(\sinh(t\m)L_p))
\nonumber\\
&& =\tr((\sinh(-t\m)L_p)L_p)  = \tr((-(\sinh(t\m)L_p))L_p)  = \tr(-\sinh(t\m)Q_p).
\nonumber
\eee

Now, decompose $F_0$:
\bee
F_0 & = & \tr \left(e^{t(\m -i\mu L_p)} Q_0\right)=F^{even}_0+F^{odd}_0,  
\eee
where
\bee
F^{even}_0 & = & \tr \left(\sum_{s=0}^{\infty}\frac 1 {(2s)!}(t(\m -i\mu L_0))^{2s} Q_0\right) =\tr \left(\sum_{s=0}^{\infty}\frac 1 {(2s)!}t^{2s}(\m^2 -\mu^2)^{s} Q_0\right)\,,                %
\label{even}
\\
F^{odd}_0&=& \tr \left(\sum_{s=0}^{\infty}\frac 1 {(2s+1)!}(t(\m -i\mu L_0))^{2s+1} Q_0\right)  
\nonumber\\
&=&\tr \left(\sum_{s=0}^{\infty}\frac 1 {(2s+1)!}t^{2s+1}(\m^2 -\mu^2_0)^{s}
(\m -i\mu_0 L_0)Q_0\right)
\nonumber\\
&=& \tr \left(\sum_{s=0}^{\infty}\frac 1 {(2s+1)!}t^{2s+1}(\m^2 -\mu_0^2)^{s}
( -i\mu_0 L_0)Q_0\right) 
\nonumber\\
&=& \sum_{s=0}^{\infty}\frac 1 {(2s+1)!}t^{2s+1}(-\mu_0^2)^{s}
( -i\mu_0)\tr L_0 
\nonumber\\
&=& \sinh (-it\mu_0)\tr L_0 = -\frac 1 {2}(z^{\mu_0}-z^{-\mu_0}) \tr L_0     .                 %
\label{odd}
\eee
%
%QQQ
%
Eq. (\ref{Fform}) implies that
\be\label{odd2}
F_{odd}= \frac 1 2
\left(  \sum_{\ell=\mu}^{-\mu}\alpha_{\ell}^{0}z^{\ell} - \sum_{\ell=\mu}^{-\mu}\alpha_{-\ell}^{0}z^{\ell}
\right).
\ee
Comparing Eq. (\ref{odd2})  with Eq. (\ref{odd}) we see that %implies
\bee                     \nonumber
\alpha^0_{\ell}&=&\alpha^0_{-\ell}, \qquad \text{ if $\ell \ne \mu$, $\ell\ne -\mu$},  \\
\alpha^0_{\mu}&-&\alpha^0_{-\mu}= -\tr L_0,
\eee
and
\be\label{127}
F_{even}=\frac 1 {2}\alpha^0_{\mu}(y^{\mu}+y^{-\mu})
+\frac 1 {2} \sum_{\ell=0}^{\mu-1} \alpha^0_{\ell}(y^{\ell}+y^{-\ell})=
 \alpha^0_{\mu}\cosh(it\mu)
+ \sum_{\ell=0}^{\mu-1} \alpha^0_{\ell}\cosh (it\ell) .
\ee

\medskip

{\proposition
\be
%\label{128}
\nonumber
\widetilde F^{\tr}_o(t) =
\alpha^0_{\mu}
+  \sum_{\ell=0}^{\mu-1} \alpha^0_{\ell}\cosh \left(t\sqrt{\mu^2-\ell^2}\,\right) .
\ee
}

\medskip

\medskip

\begin{proof}
Taking Proposition \ref {p121}  into account let us expand %decompose
Eq (\ref{121}) into the Taylor series:
\be            \nonumber
\widetilde F^{\tr}_0(t)=\sum_{s=0}^{\infty}a_{2s}\frac {t^{2s}}{(2s)!},
\ee
where $a_{2s}:=\tr(\m^{2s}Q_0)$ for $s=0,1,2,\,...$.

Eq (\ref{even}) implies
\be                 \nonumber
a_{2s}=\left( \frac {d^2}{dt^2} + \mu^2\right)^s F_{even}|_{t=0},
\ee
and Eq (\ref{127}) implies
\be                   \nonumber
a_{2s}=\left\{
\begin{array}{c}
a_{\mu }^{0} +
 \sum_{\ell=0}^{\mu-1} \alpha^0_{\ell}
\text{ \ \ \ if }s=0 \\
 \sum_{\ell=0}^{\mu-1} \alpha^0_{\ell}(-\ell^2+\mu^2)^s
 \text{ \ \ \  if }s\ne0    .
\end{array}%
\right.
\ee
So
%Further,
\be                       \nonumber
\widetilde F^{\tr}_0(t)=\sum_{s=0}^{\infty}a_{2s}\frac {t^{2s}}{(2s)!}=
\alpha^0_{\mu}
+  \sum_{\ell=0}^{\mu-1} \alpha^0_{\ell}\cosh \left(t\sqrt{\mu^2-\ell^2}\,\right)  .
\ee
\end{proof}

\subsection{Expressions for $\widetilde F^{\tr}$=$\tr\left(\exp(t\m )Q_0 \right)$ and $\widetilde F^{\tr}$ = $\tr\left(\exp(t\m ) Q_m \right)$
for the degenerate trace }

For $F_0^{(+)}$, $F_0^{(-)}$, $F_m^{(+)}$ and $F_m^{(-)}$,
we find the following expressions:
\begin{eqnarray}
F_{0}^{(+)}(z) & = & \frac{1}{i} \frac{d}{dt}\left(
\sum_{s=-d_{0}}^{s=-1}z^{-\mu _{0}-sn}\delta _{\mu _{0}>n}\right) -\frac{1}{i
}\frac{d}{dt}\left( \sum_{s=0}^{d_{0}}z^{\mu _{0}-sn}\delta _{p=0}\right)
\nonumber\\
& = & -\frac{1}{i}\frac{d}{dt}z^{\mu _{0}}+\frac{1}{i}\frac{d}{dt}\left(
-\sum_{s=1}^{d_{0}}z^{\mu _{0}-sn}\delta _{\mu
_{0}>n}+\sum_{s=-d_{0}}^{s=-1}z^{-\mu _{0}-sn}\delta _{\mu _{0}>n}\right)
\nonumber\\
& = & - \frac{1}{i}\frac{d}{dt}z^{\mu _{0}}+\frac{1}{i}\frac{d}{dt}\left(
-\sum_{s=1}^{d_0}z^{ \mu_0 - s n }\delta _{\mu_0 > n } + \sum_{s=1}^{ s = d_0 } z^{ - \mu_0 + s n } \delta_{ \mu_0 > n }\right)
\nonumber\\
& = & - \frac{1}{i}\frac{d}{dt}z^{\mu _{0}}+\frac{2}{i}\frac{d}{dt}\left(
\sum_{s=1}^{d_{0}}\sinh (it(-\mu _0 + sn))\delta _{\mu _{0}>n}\right)\,,
\end{eqnarray}

\begin{eqnarray}
F_{0}^{(-)}(z) &=&\frac{1}{i}\frac{d}{dt}\left(
\sum_{s=-d_{1}}^{s=-1}z^{-\mu _{1}-sn-m}\delta _{\mu _{1}>n}+z^{-\mu
_{1}+(d_{1}+1)n-m}\mbox{ }\delta _{n-1\geqslant m\geqslant n-\widetilde{\mu
_{1}}}\right)
\nonumber \\
&&-\frac{1}{i}\frac{d}{dt}%
\left( \sum_{s=0}^{d_{1}-1}z^{\mu _{1}-sn-m}\delta _{\mu _{1}>n}+z^{\mu_{1}- d_{1}n-m} \delta _{1\leqslant m\leqslant \widetilde{\mu }_{1}}\right)
\nonumber \\
&=&\frac{1}{i}\frac{d}{dt}\left( \sum_{s=-d_{1}}^{s=-1}z^{-\mu
_{1}-sn-m}-\sum_{s=0}^{d_{1}-1}z^{\mu _{1}-sn-m}\right) \delta _{\mu _{1}>n}
\nonumber\\
&+&
\frac{1}{i}\frac{d}{dt}\left( z^{-\mu _{1}+(d_{1}+1)n-m}-z^{\mu
_{1}-d_{1}n-m}\right) \delta _{1\leqslant m\leqslant \widetilde{\mu }_{1}}
\nonumber \\
&=&\frac{1}{i}\frac{d}{dt}\left( \sum_{s=1}^{s=d_{1}}z^{-\mu
_{1}+sn-m}-\sum_{s=1}^{d_{1}}z^{\mu _{1}-sn+n-m}\right) \delta _{\mu _{1}>n}
\nonumber\\
& + &
\frac{1}{i}\frac{d}{dt}\left( z^{-\mu _{1}+d_{1}n+m}-z^{\mu
_{1}-d_{1}n-m}\right) \delta _{1\leqslant m\leqslant \widetilde{\mu }_{1}}
\nonumber \\
&=&\frac{2}{i}\frac{d}{dt}\left( \sum_{s=1}^{s=d_{1}}\sinh (it(-\mu
_{1}+sn-m))\right) \delta _{\mu _{1}>n}
\nonumber\\
& + & \frac{2}{i}\frac{d}{dt}\sinh
(it(-\mu _{1}+d_{1}n+m))\delta _{m\leqslant \widetilde{\mu }_{1}}\,,
\end{eqnarray}

\begin{eqnarray}
F_{m}^{(+)}(z) &=&\frac{1}{i}\frac{d}{dt}\left(
\sum_{s=-d_{0}}^{s=-1}z^{-\mu _{0}-sn-m}\delta _{\mu _{0}>n}+z^{-\mu
_{0}+(d_{0}+1)n-m}\mbox{ }\delta _{n-1\geqslant m\geqslant n-\widetilde{\mu
_{0}}}\right)
\nonumber  \\
&-&\frac{1}{i}\frac{d}{dt}\left( \sum_{s=0}^{d_{0}-1}z^{\mu _{0}-sn-m}\delta
_{\mu _{0}>n}+z^{\mu _{0}-d_{0}n-m}\delta _{1\leqslant m\leqslant \widetilde{
\mu _{0}}}\right) =
\nonumber \\
&=&\frac{1}{i}\frac{d}{dt}\sum_{s=1}^{s=d_{0}}\left( z^{-\mu
_{0}+sn-m}-z^{\mu _{0}-sn+m}\right) \delta _{\mu _{0}>n}
\nonumber\\
& + & \frac{1}{i}\frac{d}{
dt}\left( z^{-\mu _{0}+d_{0}n+m}-z^{\mu _{0}-d_{0}n-m}\right) \delta
_{1\leqslant m\leqslant \widetilde{\mu _{0}}}
\nonumber \\
&=&\frac{2}{i}\frac{d}{dt}\sum_{s=1}^{s=d_{0}}\sinh (it\left( \mu
_{0}-sn+m))\right) \delta _{\mu _{0}>n}
\nonumber\\
& + & \frac{2}{i}\frac{d}{dt}\sinh
(it(-\mu _{0}+d_{0}n+m))\delta _{ m\leqslant \widetilde{\mu _{0}}}\,,
\end{eqnarray}

\begin{eqnarray}
F_{m}^{(-)}(z)&=&\frac{1}{i}\frac{d}{dt}\left( \sum_{s=-d_{1}}^{s=-1}z^{-\mu
_{1}-sn}\delta _{\mu _{1}>n}\right)
-\frac{1}{i}\frac{d}{dt}
\left( \sum_{s=0}^{d_{1}}z^{\mu _{1}-sn}\right)
\nonumber \\
&=&-\frac{1}{i}\frac{d}{dt}z^{\mu _{1}}+\frac{1}{i}\frac{d}{dt}\left(
\sum_{s=d_{1}}^{s=1}z^{-\mu _{1}+sn}\delta _{\mu _{1}>n}\right)
\nonumber\\
& - & \frac{1}{i}\frac{d}{dt}
\left( \sum_{s=1}^{d_{1}}z^{\mu _{1}-sn}\right) \delta _{\mu _{1}>n}
\nonumber \\
&=&-\frac{1}{i}\frac{d}{dt}z^{\mu _{1}}+\frac{2}{i}\frac{d}{dt}
\sum_{s=1}^{d_{1}}\sinh (it\left( -\mu _{1}+sn)\right)
\delta _{\mu _{1}>n}.
\end{eqnarray}

\vskip 5mm

\section*{Acknowledgments}
The authors
are grateful to the Russian Fund for Basic Research
(grant No.~${\text{20-02-00193}}$)
for partial support of this work.

\newpage
%%%%%%%%%%%%%%%%%%%%%%%%%%%%%%%%%%%%%%%%%%%%%%%%%%%%%%%%%%%%%%%%%%%%

\end{document}